\documentclass[graybox,natbib]{svmult}

\usepackage{txfonts}
\usepackage{lscape}
\usepackage{mathptmx}       % selects Times Roman as basic font
\usepackage{helvet}         % selects Helvetica as sans-serif font
\usepackage{courier}        % selects Courier as typewriter font
\usepackage{type1cm}        % activate if the above 3 fonts are
\usepackage{amssymb}                            % not available on your system
\usepackage{gensymb}
\usepackage{makeidx}         % allows index generation
\usepackage{graphicx}        % standard LaTeX graphics tool
\usepackage{multicol}        % used for the two-column index
\usepackage[bottom]{footmisc}% places footnotes at page bottom

\makeindex             % used for the subject index
\begin{document}

\title*{Ultra-deep Imaging: Structure of Disks and Haloes }
\author{Johan H. Knapen and Ignacio Trujillo}
\institute{Johan H. Knapen and Ignacio Trujillo \at Instituto de Astrof\'\i sica de Canarias, E-38200 La Laguna, Spain and Departamento de Astrof\'\i sica, Universidad de La Laguna, E-38206 La Laguna, Spain, \email{jhk@iac.es}
 and \email{trujillo@iac.es}}

\maketitle

\abstract*{The same as below}

\abstract{Deep imaging is a fundamental tool in the study of the outermost structures of galaxies. We review recent developments in ultra-deep imaging of galaxy disks and haloes, highlighting the technical advances as well as the challenges, and summarizing observational results in the context of modern theory and simulations. The deepest modern galaxy imaging comes from three main sources: (1) surveys such as the Sloan Digital Sky Survey's Stripe~82 project, (2) very long exposures on small telescopes, including by amateurs, and (3) long exposures on the largest professional telescopes. The technical challenges faced are common in all these approaches, and include the treatment of light scattered by atmosphere and telescope/instrument, correct flat fielding, and the subtraction of non-galaxy light in the images. We review scientific results on galaxy disks and haloes obtained with deep imaging, including the detection and characterization of stellar haloes, tidal features and stellar streams, disk truncations, and thick disks. The area of ultra-deep imaging is still very much unexplored territory, and future work in this area promises significant advances in our understanding of galaxy formation and evolution. 
}

\section{Introduction}
\label{sec:1}

Our knowledge of the properties of disks of galaxies has been driven by deep imaging for the past decades.
Until the mid-1980's, such imaging was done with photographic plates, prepared using specialist chemical
techniques and then exposed for long times on large telescopes, often by observers spending long and
uncomfortable hours in the prime focus cage of the telescope. The early history of this, and the
subsequent physical parameters derived for disk galaxies, have been summarized by, e.g.,
\citet{vdkruit:2011}. Main findings relating to the structure of disks include the description of the surface brightness distribution of disks as exponential by \citet{Freeman:1970}, and the realization that the vertical profiles of disks seen edge-on can be described by an isothermal sheet (\citealt{vdkruit:1981}). Although in the early days they were limited by a small field-of-view (FOV) and flat fielding issues, imaging with charge-coupled devices (CCDs) quickly took over from photographic plates. Large imaging surveys now provide most data, as reviewed below.

A powerful alternative to deep imaging of integrated light from galaxies is imaging and characterizing
individual stars in the outskirts of a galaxy, either by using a camera with a large FOV to observe Local Group\index{Local Group} galaxies like M31, or by using the {\it Hubble Space Telescope} ({\it HST}) to observe dwarf galaxies in the Local Group or to resolve the
stars in galaxies outside the Local Group but at distances smaller than 16\,Mpc (\citealt{Zackrisson:2012}). Exploration in this sense of M31 started with the Isaac Newton Telescope Wide Field Camera survey (\citealt{ibata:2001}) and has been recently reviewed by \citet{ferguson:2016}. Key references for other nearby galaxies include \citet{dalcanton:2009,radburn:2011,gallart:2015} and \citet{monachesi:2016}. In the current Chapter, we will concentrate on imaging of integrated light, and refer the interested reader to the Chapter by Crnojevic (this volume) for a review of results based on imaging individual stars.
 
Over the past decade, advances in detector technology, observing strategies, and data reduction procedures
have led to the emergence of new lines of research based on what we call here ultra-deep imaging. This can be obtained in large imaging surveys, or obtained for small samples of galaxies, or for individual ones, using very long exposure times on small
telescopes, or with large professional telescopes. In this Chapter, we will give examples of the first and third category, whereas impressive examples of results from the second category can be found in the Chapter by Abraham et al (this volume).

In this Chapter, we will first discuss the various challenges that need to be overcome before ultra-deep
imaging can be used to distill scientific advances, in particular those due to light scattered in the
atmosphere and telescope, flat fielding, and sky subtraction. We will then briefly review the main
approaches to obtain deep imaging, namely from imaging surveys, using small telescopes, and using large
professional telescopes. In Sect.~\ref{sec:4}, which forms the heart of this review, we consider the progress in our understanding of the outskirts of galaxies that has been achieved thanks to ultra-deep imaging, paying particular attention to the
properties of galaxy disks (including thick disks and disk truncations) and stellar haloes, but also
touching on properties of tidal streams and satellites. When concluding, we will describe future
developments, challenges and expected advances.

\section{The Challenges of Ultra-deep Imaging}
\label{sec:2}

Obtaining ultra-deep imaging of the sky is plagued with difficulties. For this reason going beyond the
30\,mag\,arcsec$^{-2}$ frontier (approximately 1500 times fainter than the darkest sky on Earth) has remained
rather elusive. In this Section, we review the most important challenges that need to be addressed carefully if one desires to
obtain ultra-deep imaging.

\subsection{Sky Brightness}
\label{sec:2sky}

Professional astronomical observatories are located on the darkest spots
on Earth. Even at these locations (where light pollution caused by human activity is minimal) the night
sky brightness\index{sky brightness} is substantial: $\mu_V\sim$22\,mag\,arcsec$^{-2}$. This brightness is mainly due to various
processes in the upper atmosphere, such as the recombination of atoms which were photoionized by the Sun
during the day, a phenomenon known as airglow\index{airglow}.  
With long enough integrations we are able to reduce the noise in this brightness which results from the intrinsic variability of the night sky,
 and easily obtain images in which we can measure features which are much fainter than the intrinsic sky brightness. However, the sky brightness also contains other components. Beyond our atmosphere, a diffuse light component is caused by the reflection of 
sunlight on the dust plane of our Solar system. This is the zodiacal light\index{zodiacal light}, with a brightness of
around $\mu_V\sim$23.5\,mag\,arcsec$^{-2}$. This brightness affects particularly those regions of the sky around
the ecliptic plane and it contaminates all observations, including those obtained with space telescopes. The intensity of
the zodiacal light is variable and depends on the Solar activity.

\subsection{Internal Reflections}
\label{sec:2int}

Internal reflections\index{internal reflections} are due to the structure of the telescope and the dome. These reflections can
appear at different surface brightness levels but are quite common when one reaches levels of $\mu_V\gtrsim$26\,mag\,arcsec$^{-2}$. There are two approaches to minimize these reflections. One way is to use telescopes
with simple optics (\citealt{abraham:2014}; see also Abraham et al, this volume), 
another is to implement clever observing strategies which avoid
repetition of similar orientation of the camera on the sky (e.g., \citealt{trujillo:2016}).

\subsection{Flat Fielding}
\label{sec:2flat}

To obtain reliable ultra-deep imaging, an exquisite flat field\index{flat fielding} correction of the images needs
to be performed. This correction is needed to ensure that a uniform illumination of the CCD leads to a uniform
output, or uniform counts in the image. In a CCD camera\index{CCD camera}, this is not the case by default as the
gain and dark current change across the face of the detector, and distortions due to optics can cause
non-uniformities. The process of flat fielding removes all these pixel-to-pixel variations in sensitivity and the
effects of distortions in the optical path. Flat field images are often obtained by exposing on uniformly
illuminated surfaces.

Any artefact (gradient, pattern, etc.) left behind during the process of creating the flat field image used to
correct the raw science images will introduce a systematic error in the final image, which in turn will prevent
reaching the expected surface brightness limit of the observation. The key to creating a good flat field image is
to have a uniform illumination of the CCD of the camera. For most observing cases, a twilight
(or even a dome) flat is good enough for this purpose. However, when the goal is to reach very faint
details of the image a different approach is needed, consisting in creating a
flat field using the night sky imaging itself. Such a flat field, using a set of science
images, is sometimes referred to as a master flat.

To create a good master flat, the set of science images must be taken at different locations on
the sky. In fact, depending on the apparent size of the galaxy, the displacement between one science
image and the next should be at least as large as the size of the object. Ideally, not all the science images of
a specific galaxy should be located at the same position on the CCD (and, ideally, should also not have the same
position angle on the sky). That means preparing an observing scheme that includes both a dithering and
a rotation pattern (see the example of this procedure in \citealt{trujillo:2016}). In addition, if the observations are
taken over different nights (or observing blocks), the best approach is to create a master flat for each
night (or observing block). The use of all the science images in a run rather than
those of a single night or observing block is not generally a good idea as slight differences from night to night in the
focus and the vignetting correction hinder such an approach.

Once all the science images have been acquired, the process of building a master flat is as follows. First, all the
objects in each individual science image are generously masked (for instance using
SExtractor; \citealt{bertin:1996}). Only the pixels outside the masked areas are used to create the final master flat.
Second, to guarantee that all the science images are appropriately weighted during their combination to create the
master flat, every individual science image of a given night is normalized. Finally, the
normalized and masked individual science images are median-combined into a single master flat.

\subsubsection{Drift Scanning}
\label{sec:2drift}

An alternative approach which has been used to achieve high-quality flat-fielding is referred to as drift
scanning\index{drift scanning}, or a variant on this called time delay and integration\index{time delay and
integration} (TDI; \citealt{mcgraw:1980,wright:1981}). In drift scanning, the reading of the CCD is done at the
same slow rate as the CCD is moved across the sky. In TDI, the CCD does not move, but the readout is timed to
coincide with the sidereal rate at which the sky passes by. In both cases, an object is sampled by every pixel in
a column, thus averaging out all defects and achieving an extremely efficient flat fielding. As in the case of the
masterflat described above, the background itself is used for flat fielding, assuring a perfect colour match.
Further details, as well as a more complete historical overview, are given by \citet{howell:2006}.

In practice, in spite of these significant advantages, drift scanning or TDI have not been used much in the
literature. The reasons for this vary, but include the difficulty of adjusting the relative movement of sky and
CCD with the readout, the loss of efficient observing during the ramp-up and ramp-down phases at the start and end
of an exposure, image elongation effects, and the fact that the exposure time is fixed by the telescope$+$CCD
setup, and often rather short. 

In the field of imaging nearby galaxies, the most notably exception to this general dearth of TDI results is the
Sloan Digital Sky Survey (SDSS, \citealt{york:2000}). As described by \citet{gunn:1998}\footnote{For a more
detailed description of the flat field procedures used by the SDSS, see
http://classic.sdss.org/dr5/algorithms/flatfield.html.}, the SDSS large-format mosaic CCD camera has been designed
to image strips of the sky simultaneously in five colour bands using the TDI technique. The approach chosen by the
SDSS team has proven to be very successful in terms of imaging to low surface brightness levels. Even though the
exposure time of SDSS images is less than one minute (53.9\,s) and the telescope of modest size (2.5\,m), the
exquisite flat fielding and sky background allow one to reach very low surface brightness levels indeed, down to
26.5\,mag\,arcsec$^{-2}$ ($3\sigma$ in an area of $10\times10$\,arcsec; \citealt{trujillo:2016}) or down to 27.5\,mag\,arcsec$^{-2}$ when analyzing elliptically averaged surface brightness profiles (see, e.g., \citealt{pohlen:2006}). As illustrated in
Sect.~\ref{sec:3}, co-adding series of SDSS images, as is possible in the Stripe~82 survey area, brings that level
down by another 2\,mag\,arcsec$^{-2}$, allowing ground-breaking science to be performed.

\subsection{Masking and Background Subtraction}
\label{sec:2mask}

\begin{figure}[ht]
%\sidecaption
% For example, with the graphicx style use
\centering
\includegraphics[width=\textwidth]{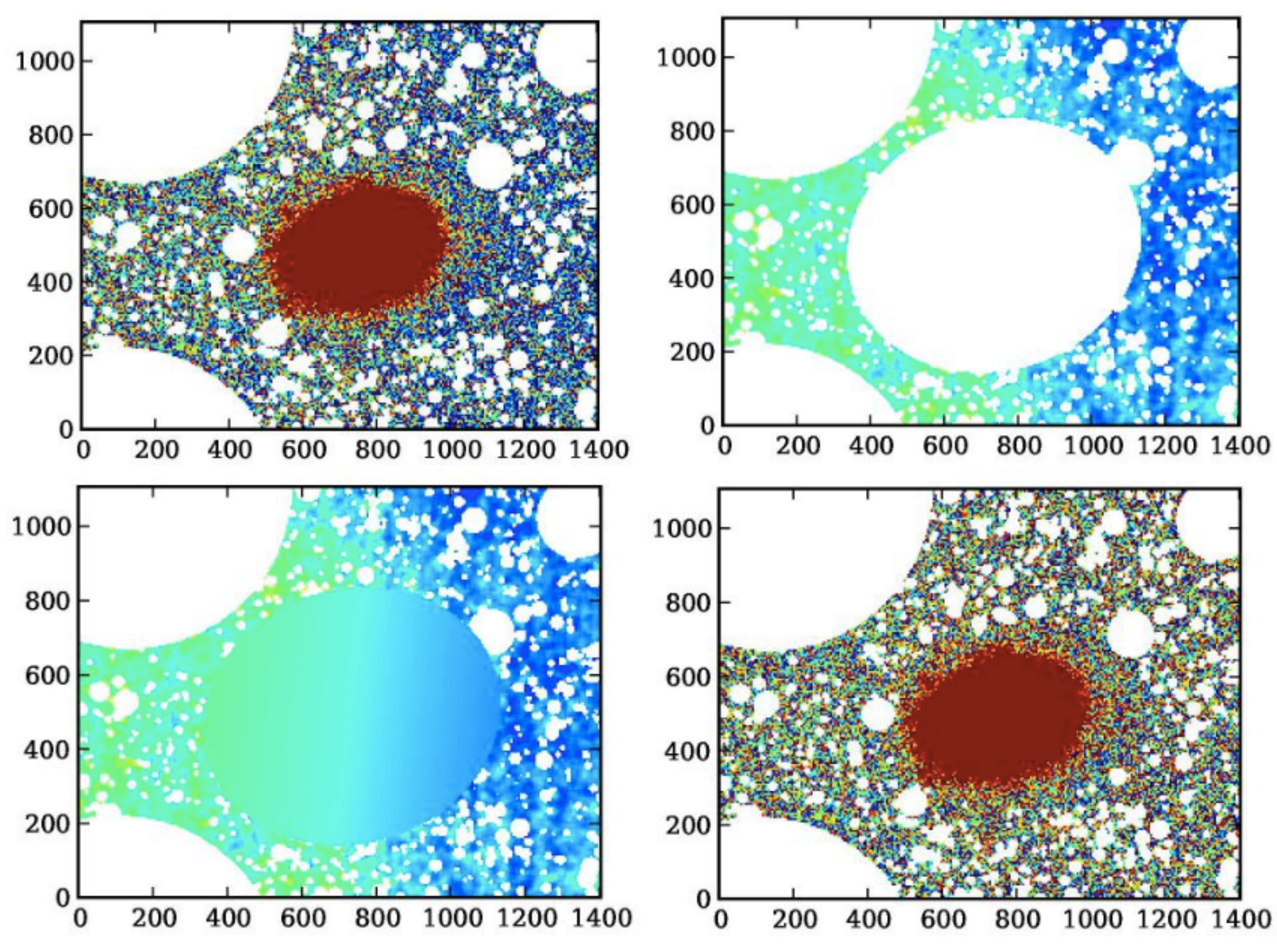}
%
% If no graphics program available, insert a blank space i.e. use
%\picplace{5cm}{2cm} % Give the correct figure height and width in cm

\caption{Example of residual background modelling and subtraction, for a Stripe~82 image of the galaxy NGC~941.
{\it Top left} panel: original image, showing a gradient in the background level. White areas are masked-out images of
foreground stars and background galaxies. Image labels are in pixels, with size 0.396\,arcsec. North is right,
East to the bottom. {\it Top right}: model for the background, excluding the area of the galaxy. {\it Lower left}: background
model, extrapolated over the area of the galaxy. {\it Lower right}: final image, with the background model subtracted
from the original. Reproduced with permission from \citet{peters:2017}}

\label{background}       % Give a unique label
\end{figure}

Deep images often reveal coherent structure at low levels, which can be due to, e.g., imperfect flat fielding,
spatial variations in the background sky, or residual point spread function (PSF) effects. Deep images of galaxies
also show foreground stars, as well as background galaxies. All these components must be identified, taken into
account, and/or subtracted before a deep galaxy image can be analysed. As an example of the procedures followed,
we show in Fig.~\ref{background} a SDSS Stripe~82 image of NGC~941 as analysed by \citet{peters:2017}. Stars and
background galaxies are generously masked, after which a polynomial two-dimensional fit is made to the remaining
background pixels. In our example, this is dominated by a left-right gradient in the background level, but the fit
also shows smaller-scale structure. The latter could, in principle, be real structure, either related to the
galaxy (e.g., tidal streams) or not (e.g., Galactic cirrus, see Sect.~\ref{cirrus}). If one looks for structure
like tidal streams a different background modelling technique must be used, for instance only modelling
large-scale fluctuations or gradients. This neatly highlights the difficult nature of this kind of analysis.

In our example, the gradient in the background in all probability is also present at the location of the galaxy,
which is why we extrapolate the background model into the galaxy region (lower left panel of
Fig.~\ref{background}). This model is then subtracted from the original image, and the result can be used for
scientific analysis, in this case studying the shape of the outer regions through analysis of azimuthally averaged
radial profiles (see Sect.~\ref{sec:4trunc}; this is why the small-scale background structure described in the
previous paragraph could safely be subtracted off in this case). The uncertainty limit, down to which these radial
profiles can be trusted, is just below 30\,mag\,arcsec$^{-2}$ for the image shown in Fig.~\ref{background}
(data taken from the IAC Stripe82 Legacy Project; \citealt{fliri:2016}).

\subsection{Scattered Light}
\label{sec:2scat}

\begin{figure}[ht]
%\sidecaption
% For example, with the graphicx style use
\centering
\includegraphics[width=0.98\textwidth]{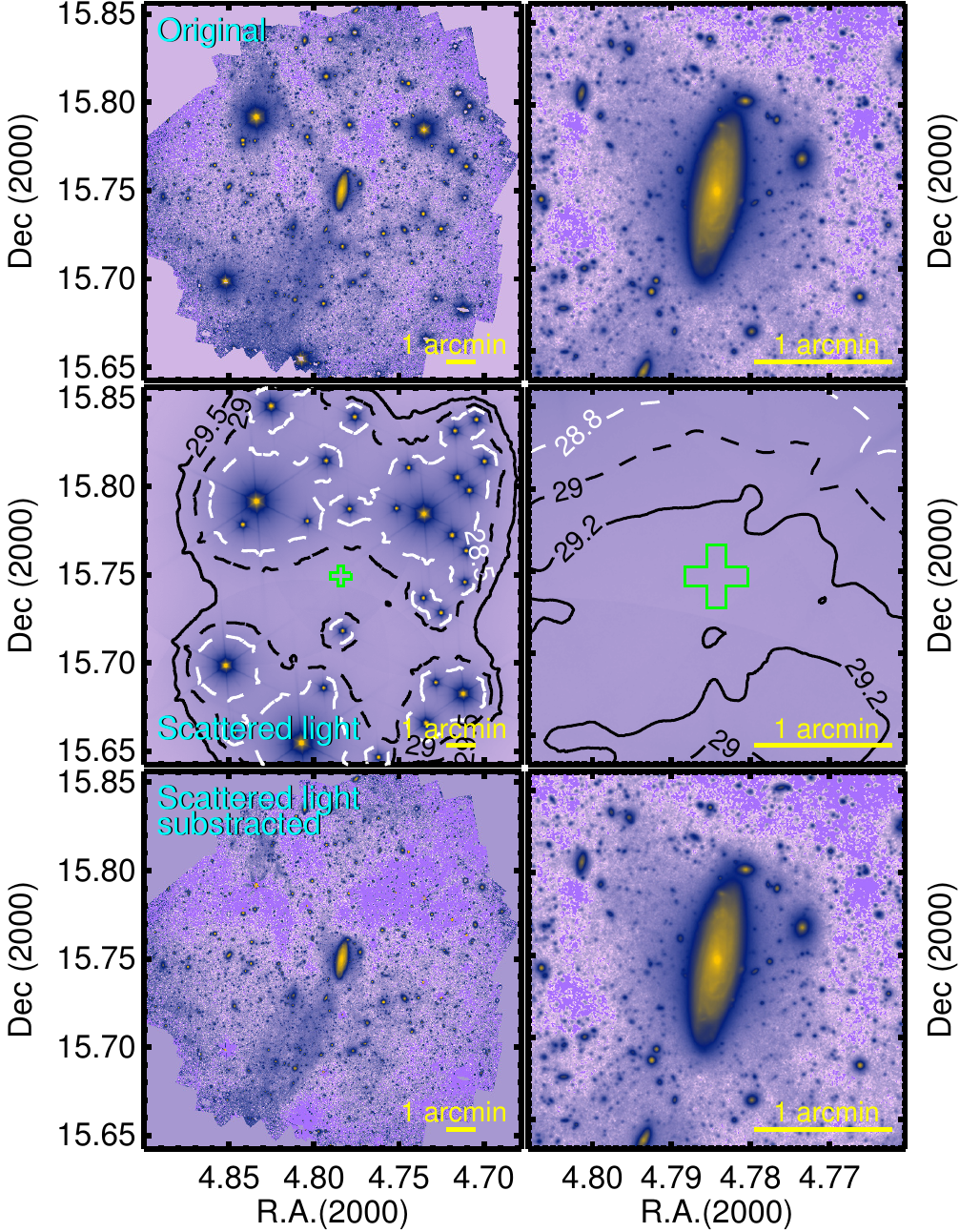}
%
% If no graphics program available, insert a blank space i.e. use
%\picplace{5cm}{2cm} % Give the correct figure height and width in cm

\caption{The scattered light around the galaxy UGC~00180 produced by all the stars brighter
than $R=17$\,mag in its vicinity. {\it Top row}: original field (left) and a zoom-in of the galaxy. {\it Middle row}: scattered light. The position of the galaxy is illustrated
with a green cross. Contours of surface brightness are 28.5, 29, and 29.5\,mag\,arcsec$^{-2}$ ({\it left}) and 28.8, 29, and 29.2\,mag\,arcsec$^{-2}$ ({\it right}). {\it Lower row}:
original field after subtraction of the scattered light produced by the brightest sources. Figure
taken from \citet{trujillo:2016}, reproduced with permission of the AAS}

\label{scatter}       % Give a unique label
\end{figure}

A further and important source of background contamination is produced by all the emitting sources in the image (or even just outside the imaged area).
The light of these sources is scattered by the \index{point spread function} PSF of the instrument across the
entire image. \citet{slater:2009} have shown that at $\mu_V\sim$29.5\,mag\,arcsec$^{-2}$, an image taken from the
ground has all its pixels affected by scattered light from nearby bright sources. Properly removing this
contamination is quite challenging, and requires an extremely accurate characterization ($<$1\% at large
distances) of the PSF of the camera.

The excess light redistributed by the PSF from both the object of interest and the nearby surrounding sources
creates a two-dimensional and  highly structured surface that is the main contributor to the background of the
image at the faintest surface brightness levels ($\mu_V\gtrsim$29\,mag\,arcsec$^{-2}$, \citealt{slater:2009}). All
astronomical images are affected by this scattered light background which is the result of the convolution of the
PSF with the light of the sources. The scattered light background of an astronomical image will be more intense if
the number of bright sources in the image is large and also if the PSF  has significant wings. In this sense, the
best option (if feasible) is to select a target within a field devoid of nearby bright surrounding objects. A
typical scattered light background is illustrated in Fig.~\ref{scatter}.

Figure \ref{scatter} illustrates how to deal with the background of scattered light. First, the PSF must be characterized as perfectly as possible by using a reliable extended PSF. An
extended PSF with high a signal-to-noise ratio in its outer wings will allow the exploration of the distribution of the
light of the nearby brightest sources up to the position of the galaxy under exploration. Second, using
the extended PSF a background scattered light map is created. If the main contributor to the scattered
light background is the presence of bright stars then the scattered light map is created by simply locating
the model PSF at the position of the bright stars and scaling the flux of the PSF to these
bright sources. Finally, the scattered light map is subtracted from the observed image. 

Above, we have discussed the effect of the scattered light created by the surrounding sources around the galaxy of
interest. Naturally, the light distribution of the object itself is also convolved with the PSF.  In this sense,
the scattered light of the targeted galaxy is also creating an artificial excess of light in the outermost region
of the galaxy that needs to be addressed in order to explore the properties of the object in its outer regions. We
illustrate how to deal with this in Sect.~\ref{sec:4halo_psf} of this Chapter.

\subsection{Galactic Cirrus}
\label{cirrus}

Finally, if one is interested in exploring the fainter structures of galaxies, the presence of Galactic
cirrus\index{Galactic cirrus} in the images needs to be considered. These filamentary structures are located everywhere on the
sky, even at higher Galactic latitudes. For this reason, a careful preselection of the fields to be
observed is necessary to minimize the contamination by cirrus. A good illustration of the perils of Galactic cirrus is presented by \citet{davies:2010} for the case of the M81 group. They found that far-infrared emission measured by the {\it Herschel} satellite correlates very well spatially with narrow-velocity Galactic H{\sc i}, without any evidence that this far-infrared emission originates in the M81 group. They thus inferred that the optical streams and structures seen in the M81 group are not in fact part of the group, but are rather due to light from our own Galaxy which is back-scattered off Galactic dust.

\section{Approaches in Ultra-deep Imaging}
\label{sec:3}

The last few years have seen an explosion in the number of works exploring the outermost regions of 
nearby galaxies using very deep imaging. These works can be grouped into three different flavours: deep
($\sim1$\,h) multipurpose surveys with medium sized ($2-4$\,m) telescopes, extremely long integrations
($\gtrsim 20$\,h) of particular galaxies with small ($\lesssim 1$\,m) telescopes or  long integrations
($\gtrsim 5$\,h) with large ($\gtrsim 8$\,m) telescopes. In what follows we will summarize some of these
efforts.

\subsection{Survey Data}
\label{sec:3survey}

The  3.6\,m Canada-France-Hawaii Telescope (CFHT), with its MegaCam wide-field camera comprising 36 CCDs and
covering a 1\,square degree FOV, has played a significant role in the new generation of  deep imaging
surveys\index{imaging surveys}. This telescope has been used for general purpose surveys like the Wide Synoptic
CFHT Legacy Survey (155 square degrees; \citealt{Cuillandre:2012}), or more specific projects like the Next
Generation Virgo Cluster Survey (NGVCS; \citealt{Ferrarese:2012}) or the deep imaging follow-up
(\citealt{Duc:2015}) of the ATLAS$^{3{\rm D}}$ project (\citealt{Cappellari:2011}). These surveys are
characterized by having similar depths ($r\sim25-25.5$\,mag; $S/N=10$ for point sources) using integration times 
of $40-60$\,min, reaching a limiting surface brightness of $28.5-29$\,mag\,arcsec$^{-2}$ ($3\sigma$ in
$10\times10$\,arcsec$^2$). 

Among the most important results of the NGVCS is the connection between the
distribution of the metal-poor globular clusters and the intra-cluster light (\citealt{Durrell:2014}), suggesting
a common origin for these two structures. The deep imaging by \citet{Duc:2015} revealed a large number of features
surrounding their galaxies. Both these projects targeted mostly early-type galaxies. Unfortunately, the
pipeline used for the reduction of the Wide Synoptic CFHT Legacy Survey removed the low surface brightness
features around the brightest extended galaxies in the images, producing obvious ``holes" that prevent the use of
this reduced dataset for the exploration of the outermost parts of the spiral galaxies. In the later MegaCam
surveys this problem does not occur.

Another telescope that has revolutionized our understanding of galaxies is the Sloan 2.5\,m telescope.
The Sloan telescope is well known for producing the SDSS (\citealt{york:2000}).
Among the different projects that Sloan has covered, of particular interest here is its deep imaging
survey in the Southern Galactic cap, popularly known as the ``Stripe~82" survey
(\citealt{jiang:2008,abazajian:2009}). The Stripe~82 survey covers an area of 275 square degrees along
the celestial equator ($-50\degree <$ R.A. $< 60\degree$, $-1.25 <$ Dec. $< 1.25$) and has been observed
in all the five SDSS filters: $ugriz$. The typical amount of time on source was $\sim$1.2\,h. Being
located at the equator, the Stripe~82\index{Stripe 82} area is accessible from most ground-based facilities.  A third
of all the available SDSS data in the Stripe~82 area were combined by \citet{annis:2014}. They reached a
depth (50\% completeness for point sources) of $r\sim24.2$\,mag. Later on, \citet{jiang:2014}
used the entire dataset and reported a gain of $0.3-0.5$\,mag in depth compared to the previous reduction.
None of these reductions were done with the aim of exploring the lowest surface brightness features of
the objects. This task was performed by \citet{fliri:2016} and is known as the IAC Stripe~82 Legacy
Project. \citet{fliri:2016} reached a depth of $r\sim24.7$\,mag (50\% completeness for point
sources) and a limiting surface brightness of $\mu_r\sim28.5$\,mag\,arcsec$^{-2}$ ($3\sigma$ in $10\times10$\,arcsec$^2$). 
The reduced images of \citet{fliri:2016} have been made publicly available through a dedicated
webpage (http://www.iac.es/proyecto/stripe82).

\subsection{Small Telescopes}
\label{sec:3small}

As surface brightness is independent of telescope aperture, in principle one can use small telescopes to reach
ultra-faint surface brightness levels. Direct advantages of using small telescopes over larger ones include the
larger field of view, and the reduced competition for observing time, in particular when private telescopes are
used. Using modern CCD technology, this has been done by several workers in the field, aiming to uncover light
sources as diverse as intra-cluster light, diffuse galaxies, or the outer regions of bright galaxies. The Chapter
by Abraham et al (this volume) gives significantly more detail on this; we review the basics in this short Section.

Mihos and collaborators used the Case Western Reserve University's Burrell Schmidt 0.6\,m telescope to obtain
ultra-deep imaging of the core region of the Virgo cluster, down to limits of $\mu_V=28.5$\,mag\,arcsec$^{-2}$ (see \citealt{mihos:2016} for a recent update).
From these images, they were able to reveal a number of interesting features of the intra-cluster
light\index{intra-cluster light}, including several tidal streamers of more than 100\,kpc in length, and many
smaller tidal tails and bridges between galaxies, and to conclude that cluster assembly appears to be hierarchical
in nature rather than the product of smooth accretion around a central galaxy (\citealt{mihos:2005}; the colours
of features in the intra-cluster light were measured by \citealt{rudick:2010}). \citet{mihos:2015} used the same
data set to find three large and extremely low surface brightness galaxies, only one of which shows the signs of
tidal damage that might be expected for such galaxies in a dense cluster environment. The same telescope was used
by \citet{mihos:2013} to image the galaxy M101 over an area of 6\,deg$^2$ and to a depth of
$\mu_B=29.5$\,mag\,arcsec$^{-2}$, and by \citet{watkins:2014} to push down to $\mu_B=30.1$\,mag\,arcsec$^{-2}$ in
the M96 galaxy group. The former authors found a number of plumes and spurs but no very extended tidal tails,
suggestive of ongoing evolution of the outer disk of the galaxy due to encounters in its group environment,
whereas the latter found no optical counterpart to the extended H{\sc i} surrounding the central elliptical
M105, and in general only very subtle interaction signatures in the M96 group.

In a fruitful collaboration with amateur astronomers,
\citet{martinez-delgado:2008,martinez-delgado:2010,martinez-delgado:2015} have used small private telescopes to
obtain deep images of tidal streams\index{tidal streams} around a number of galaxies, most with pre-existing
evidence for the presence of some kind of outer structure, and more recently of low surface brightness galaxies in
the fields of large nearby galaxies (e.g., \citealt{javanmardi:2016}). This group uses very long exposures taken
at dark sites, imaging through wide filters. Among the most beautiful and well-known results obtained by Mart\'\i
nez-Delgado's group is the discovery of the optical analogues to the morphologies predicted from $N$-body models
of stellar haloes constructed from satellite accretion (e.g., \citealt{bullock:2005,johnston:2008}). While the
resemblance between models and observations is indeed striking and important, it must be kept in mind that most of
the structure seen in the models is at lower or much lower surface brightness levels than even the deepest
currently available imaging, and that most galaxies observed by Mart\'\i nez-Delgado et al were targeted
specifically to have some previous evidence for tidal structure and are thus not representative of the general galaxy
population. In fact, many nearby galaxies show no evidence at all for any tidal or other disturbances in deep
images (e.g., \citealt{Duc:2015}, see also \citealt{merritt:2016}).

A third strand, besides using small existing research telescopes or amateur installations, is to use simple,
small, custom-built telescopes optimized for deep imaging through simple optics and a very careful treatment of
systematics. The best-known example of this is the Dragonfly Telephoto Array (\citealt{abraham:2014}), a set of up
to 48 commercial telephoto lenses with excellent coatings coupled to CCD cameras, which minimizes the amount of
scattered light produced inside the telescope. In the current configuration, the array is equivalent to a 1\,m
aperture telescope with a field of view of 6 square degrees (\citealt{abraham:2014,abraham:2016}; see also Abraham et al, this volume).

Among the results obtained with Dragonfly images and profiles (which go down to $\mu_g\sim28$\,mag\,arcsec$^{-2}$;
3$\sigma$ in $12\times12$\,arcsec boxes; \citealt{merritt:2016}) are the finding that there is a significant spread
in  the stellar mass fraction surrounding galaxies (\citealt{vandokkum:2014,merritt:2016}, see also
Sect.~\ref{sec:4halo}), and a study of a so-called ultra-diffuse galaxies in the Coma cluster
(\citealt{vandokkum:2016} and references therein). 

\subsection{Large Telescopes}
\label{sec:3large}

Large (8-10\,m class) telescopes have barely been used so far to obtain very deep imaging of nearby galaxies. To
the best of our knowledge the first successful attempt of going ultra-deep (i.e., surpassing the
30\,mag\,arcsec$^{-2}$ barrier)  was conducted by \citet{jablonka:2010}, who used the imaging mode of the VIsible
MultiObject Spectrograph (VIMOS) on ESO's Very Large Telescope (VLT) to target the edge-on S0 galaxy NGC~3957
reaching surface brightness limits (Vega system) of $\mu_R=30.6$\,mag\,arcsec$^{-2}$ (1$\sigma$; 6\,h) and 
$\mu_V=31.4$\,mag\,arcsec$^{-2}$  (1$\sigma$; 7\,h). These authors found that the stellar halo of this galaxy,
calculated between 5 and 8 kpc above the disk plane, is consistent with an old and preferentially metal-poor
normal stellar population, like that revealed in nearby galaxy haloes from studies of their resolved stellar
content. Also worth mentioning is the work by \citet{galaz:2015},  who used the more ``modest" 6.5\,m Magellan
telescope to observe the extremely large galaxy Malin~1. \citet{galaz:2015} used the Megacam camera to image the
galaxy for about 4.5\,h in $g$ and $r$, reaching $\mu_B\sim$28\,mag\,arcsec$^{-2}$. Using these images, they
obtain an impressive result for the diameter of Malin 1 of 160 kpc, $\sim$50 kpc larger than previous estimates.
Their analysis shows that the observed spiral arms reach very low luminosity and mass surface densities, to
levels much lower than the corresponding values for the Milky Way.

\begin{figure}[htb]
%\sidecaption
% For example, with the graphicx style use
\includegraphics[width=\textwidth]{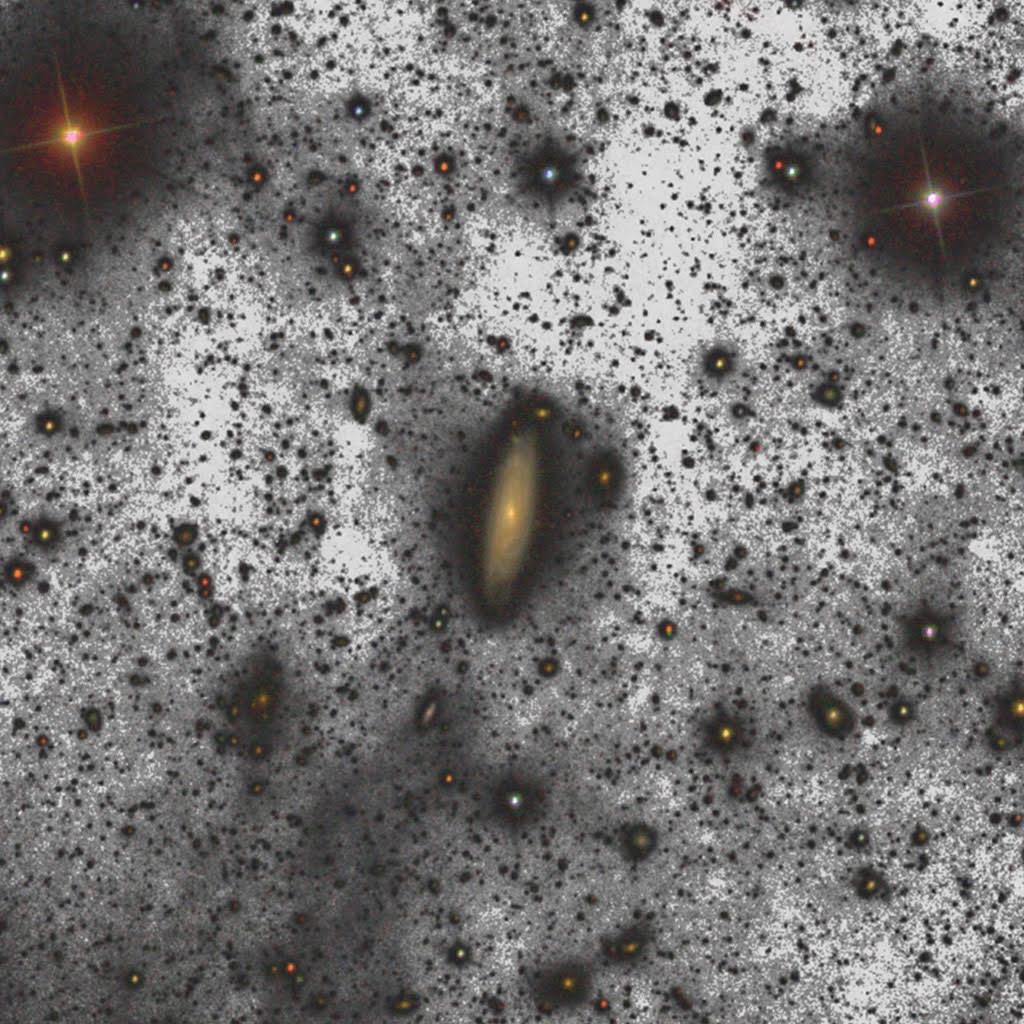}
%
% If no graphics program available, insert a blank space i.e. use
%\picplace{5cm}{2cm} % Give the correct figure height and width in cm

\caption{The galaxy UGC~00180 and its surrounding field observed down to a surface brightness limit of
31.5\,mag\,arcsec$^{-2}$ in the $r$-band (around 10 times deeper than most of the previous deep images obtained from the
ground). The image is a combination of SDSS imaging (colour part) and 8.1\,h of imaging with the GTC; grey
part). In addition to the stellar halo of the galaxy, the image shows a plethora of details. Among the
most remarkable is the filamentary emission from dust of our own Galaxy located in the bottom-left part
of the image. There are also distant galaxies which are seen to be merging with other objects, and a
high-redshift cluster towards the bottom-left corner of the galaxy where the intra cluster light is visible.
Credit: GTC, Gabriel P\'erez and Ignacio Trujillo (IAC)}

\label{fig:1}       % Give a unique label
\end{figure}

The currently deepest ever image of a nearby galaxy was recently obtained by \citet{trujillo:2016}. These
authors pointed the Gran Telescopio Canarias (GTC, a 10.4\,m telescope) at the galaxy UGC~00180, an object similar to M31 but located at a
distance of $\sim150$\,Mpc (see Fig.~\ref{fig:1}). Their $r$-band image  reached a limiting surface brightness of
31.5\,mag\,arcsec$^{-2}$ (3$\sigma$; $10\times10$\,arcsec$^2$)  after 8.1\,h on-source integration. This image
revealed a stellar halo with significant structure surrounding the galaxy. The stellar halo has a mass fraction
of $\sim$3\% of the total stellar mass of the galaxy. This value is close to the one found in the Milky Way and M31 using
star counting techniques. This is the first time that integrated-light observations of galaxies reach a  surface
brightness depth close to that achieved using star counting techniques in nearby galaxies. It is a major step
forward as it allows the exploration of the stellar haloes in hundreds of galaxies beyond the Local Group, in
particular when viewed in the context of future imaging possibilities with the Large Synoptic Survey Telescope
(LSST; see Sect.~\ref{sec:5}).

\section{Disk and Stellar Halo Properties from Ultra-deep Imaging}
\label{sec:4}

\subsection{Thick Disks}
\label{sec:4thick}

A thick disk is seen as an excess of flux above and below the mid-plane of edge-on and highly inclined galaxies,
typically at a few times the scale height of the thin disk component and with a larger exponential scale height
than the thin disk which is dominant in the mid-plane regions. Thick disks\index{thick disks} in external galaxies
were discovered and defined by \citet{tsikoudi:1979}  and \citet{burstein:1979}, and have since been found to be
very common, in all kinds of galaxies (see, e.g., \citealt{dalcanton:2002}, \citealt{yoachim:2006}, \citealt{comeron:2011b},
\citealt{comeron:2011a}). Our Milky Way has also been known for a long time to have a thick disk component (e.g.,
\citealt{yoshii:1982}, \citealt{gilmore:1983}), with stars in the thick disk having significantly lower
metallicity than those in the thin component (\citealt{gilmore:1985}).

There is an ongoing debate on the issue of the definition of a thick disk. Early work on the thick disk in the
Milky Way, and practically all work on extragalactic thick disks, basically fits stellar density distributions
with thin and thick disk components, thus defining them {\it geometrically}. It is natural to assume that the
stars forming part of the thus defined thick component have higher velocity dispersion, and possibly older ages
and lower metallicity, than those in the thin disk. In our Milky Way, one can observe individual stars and one can
thus define the thick disk {\it chemically}, for instance by measuring their $\alpha$-element abundances relative
to iron. Some of the interesting consequences of this are the confirmation by \citet{bensby:2014} that the scale
length of the thick disk is much shorter than that of the thin disk, or even the statement by
\citet{bovy:2012} that the Milky Way does not have a distinct thick disk at all but rather a continuous
distribution of disk thicknesses (but see comments by \citealt{haywood:2013}). This is an ongoing discussion which
we note but will not review here. The interested reader may find more discussion and references in a recent
conference discussion published by \citet{kawata:2016}.

Different formation mechanisms, which can be grouped into three main families, have been proposed for thick disks.
All mechanisms directly link the properties of thick disks to the evolution of galaxies, and some indicate clear
constraints on the formation of disks. The latter category includes the first broad class of models, in which the
high velocity dispersion of the material forming the disk at high redshift led to the thick disk; the thin disk
formed later from lower-dispersion material (e.g.,
\citealt{samland:2003,brook:2004,elmegreen:2006,bournaud:2009}). The second class of models implies a secular
origin for thick disks, by stipulating that they are caused by vertical heating\index{vertical heating} and/or the
radial migration\index{radial migration} of stars (e.g., \citealt{villumsen:1985,schonrich:2009}; see also the
review by Debattista et al, this volume). In the third class of models the thick disk arises from interactions with
satellite galaxies\index{satellite galaxies}, through the accretion\index{accretion} of stars (e.g.,
\citealt{quinn:1993}) or through dynamical heating (e.g., \citealt{abadi:2003}). No consensus exists as yet in the
literature as to which model is the most adequate, and it is likely that all play some role in the formation of
thick disks. 

Focussing now on the detection and characterisation of extragalactic thick disks, there are two main techniques in
use: star counts\index{star counts} and direct imaging. The former can be done only in the nearest of external
galaxies and only with the {\it HST}, the latter, in principle, across much larger samples (see below). Older work
on resolved stars with {\it HST} includes that by \citet{mould:2005}, \citet{tikhonov:2005}, \citet{seth:2005},
and \citet{rejkuba:2009}, who use resolved red giant branch (RGB) stars to characterize the thick disk component
in one or a handful of galaxies. Recent {\it HST} work includes that by \citet{streich:2016} who used images from the
GHOSTS survey to note the absence of a separate  thick disk component for three of their survey galaxies (although
their choice of galaxies was not optimal: they are of low mass, so dust lanes are weak or absent and hence the
inclination cannot be established well---an important point as thick disks are optimally studied in edge-on
galaxies).

Among the most comprehensive studies of samples of external galaxies is that by \citet{yoachim:2006} who used
mainly $R$-band imaging of 34 late-type edge-on disk galaxies to measure the main structural parameters of the
thin and thick disk components by fitting one-dimensional analytic expressions based on the generalized function
sech$^{2/n}$. Although such functions give decent fits to profiles, they are essentially {\it ad hoc} and, in the
case of superposed fits of both a thin and a thick disk, ignore the gravitational interaction between the two disk
components and therefore are not well justified physically.

\begin{figure}[htb]
%\sidecaption
% For example, with the graphicx style use
\includegraphics[width=\textwidth]{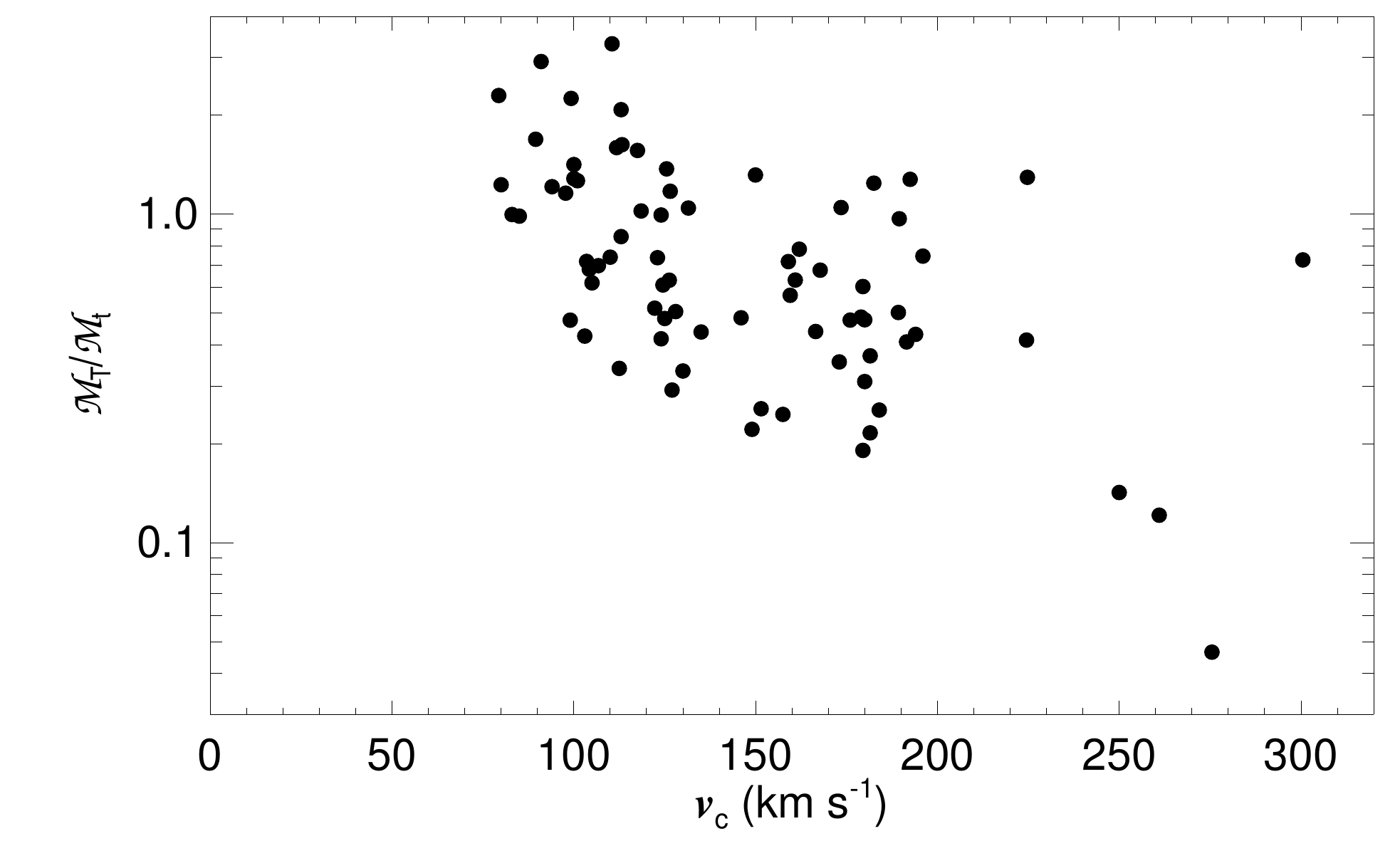}
%
% If no graphics program available, insert a blank space i.e. use
%\picplace{5cm}{2cm} % Give the correct figure height and width in cm

\caption{Ratio of the mass of the thick to that of the thin disk, $M_{\rm T}/M_{\rm t}$, as function of circular velocity $v_{\rm c}$ for a sample of edge-on galaxies expanded from that of \citet{comeron:2012}. The new analysis underlying these data incorporates PSF modelling and correction. A value of unity indicates equal thick and thin disk masses. Figure reproduced with permission from S. Comer\'on et al (in prep.)}

\label{comeron2012}       % Give a unique label
\end{figure}

To remedy this, \citet{comeron:2011b} integrate the equations of equilibrium for a set of gravitationally coupled
isothermal stellar and gas disks to derive the parameters of the thin and thick disk components, in particular
their relative masses. These authors used deep imaging at 3.6\,$\mu$m from the {\it Spitzer} Survey of Stellar
Structure in Galaxies (S$^4$G; \citealt{sheth:2010}), which have the advantage of being essentially unaffected by
dust---an important consideration when studying edge-on galaxies\index{edge-on galaxies}. Fitting one-dimensional
luminosity profiles obtained from the images of 46 edge-on and highly inclined galaxies, after a careful masking
and background-modelling and -subtraction effort, \citet{comeron:2011b} found that thick disks are not only
ubiquitous, but also significantly more massive than previously reported. Typically, thick and thin disks have comparable masses
(Fig.~\ref{comeron2012}), which favours an {\it in situ} origin for the thick disk component, with possibly
significant additional amounts of stars from satellites that were accreted after the formation of the galaxy, or
from secular heating of the thin disk. Because of the different mass-to-light ratios in thin and thick disks, the
reported higher mass fractions in the thick component lead to higher overall disk masses, and thus a reduced need
for dark matter. 

In a later paper, \citet{comeron:2012} confirmed this basic result with a somewhat larger sample of 70 edge-on
galaxies, while \citet{comeron:2014} used an analysis of the thin and thick disk components as well as of the
central mass concentrations (CMCs; ``bulges") to conclude that the ratio of the mass of the dynamically hot
components (thick disk and CMC) and that of the cold components (thin and gas disks) is constant and does not
depend on the mass of the galaxy. This suggests that both the thick disk and the CMC were formed early on, in a
short phase of intense star formation activity, and in a turbulent gas disk. The other components were formed
later, in a slower phase of lower star formation intensity. Recent work (S. Comer\'on et al, in prep.) shows that
PSF effects in the S$^4$G imaging are small enough not to influence any of these main conclusions. 

Future work will focus on a number of different areas. Firstly, building larger samples of galaxies with
accurately determined thick disk properties, both locally and at higher redshift. The deepest {\it HST} imaging
(e.g., of the {\it Hubble} ultra-deep fields) can in principle be used for the latter, while the LSST (see
Sect.~\ref{sec:5}) should provide the deep imaging of huge samples of nearby galaxies. Secondly, obtaining
detailed information on the kinematics and stellar populations of the thick disks. This is starting to be done now
with large telescopes. For instance, \citet{comeron:2015}  used observations with the VIMOS instrument on ESO's
Very Large Telescope (VLT) to deduce from its kinematics and stellar populations that the thick disk in the galaxy
ESO~533-4 formed in a relatively short event, \citet{comeron:2016} used MUSE on the VLT to conclude that the thick
disk in the S0 galaxy ESO~243-49 formed early on in the history of the galaxy, and before the star formation in
the galaxy was quenched, while \citet{kasparova:2016} and \citet{guerou:2016} provide further evidence that the
formation mechanisms of thick disks in galaxies are diverse. Deep spectroscopy with large telescopes, as well as
careful colour measurements from LSST imaging, will no doubt bring further progress here, and thus shine light on
the formation of disk galaxies.

\subsection{Truncations} \label{sec:4trunc}

The relatively sharp edges of edge-on stellar disks, referred to as truncations\index{truncations}, were first
noted by \citet{vdkruit:1979}. Such truncations, typically occurring at radii of around four or five times the
exponential scale-length of the inner disk and with scale lengths of less than 1\,kpc, appear to be very common,
with about three out of four thin discs truncated (\citealt{comeron:2012}, see the review by
\citealt{vdkruit:2011} for further details). 

Truncations are important as they are potential key indicators of the formation and early evolution processes that
have shaped disk galaxies, as reviewed in detail by \citet{vdkruit:2011}, in particular in their Sect.~3.8. For
instance, as the truncation corresponds to the maximum in the distribution of specific angular
momentum\index{distribution of specific angular momentum} across the disk, it may indicate directly what that
distribution was in the proto-galaxy, in the case of conservation of angular momentum. Re-distribution of material
in the disk, as reviewed by Debattista et al (this volume), may have occurred, leading to the
current distribution of angular momentum being unrelated to that of the material that formed the disk. In
addition, as reviewed by Elmegreen and Hunter (this volume), disk breaks\index{disk breaks} and truncations are
intimately related to the past and present star formation processes in disks. 

While it is thus important to characterize truncations, only in edge-on galaxies\index{edge-on galaxies} are they
bright enough to have been observed routinely for the past decades, first on the basis of photographic imaging and
later using CCDs. Imaging them in face-on or moderately inclined galaxies has so far proven mostly elusive. Not
only will truncations occur at much fainter levels there because of reduced line-of-sight integration through the
disk, they can also be masked by the lopsided nature of spiral galaxies (e.g., \citealt{zaritsky:2013}), or confused
by the onset of stellar haloes (\citealt{martin-navarro:2014}). In addition, ``disk" breaks at relatively high
surface brightnesses and occuring at $\sim8\pm1$\,kpc in edge-on galaxies have been confused in the literature
with truncations, at $\sim14\pm2$\,kpc (\citealt{martin-navarro:2012}). For instance, the features labelled as
truncations by \citet{pohlen:2006} most probably are disk breaks rather than the face-on counterparts of the
truncations observed in edge-on galaxies. 

Using some of the deepest imaging available for samples of nearby galaxies, the SDSS Stripe~82 dataset (see
\citealt{fliri:2016} and Sect.~\ref{sec:3survey}), \citet{peters:2017} have recently attempted to find the elusive
truncations in a sample of 22 face-on to moderately inclined galaxies. They used the new data products from
\citet{fliri:2016} and added the $g', r'$ and $i'$ images to reach extra depth. They selected their galaxies to be
undisturbed, and with well-behaved image backgrounds, and were able to extract surface photometry down to
$29-30\,r'$-mag\,arcsec$^{-2}$ after performing careful but aggressive masking and modelling of residual
background gradients (see Fig.~\ref{background} for an example). \citet{peters:2017} then used and compared a
variety of different analysis and extraction methods to optimize the detection and characterisation of truncations
in their sample galaxies.

\begin{figure}[htb]
%\sidecaption
% For example, with the graphicx style use
\includegraphics[width=\textwidth]{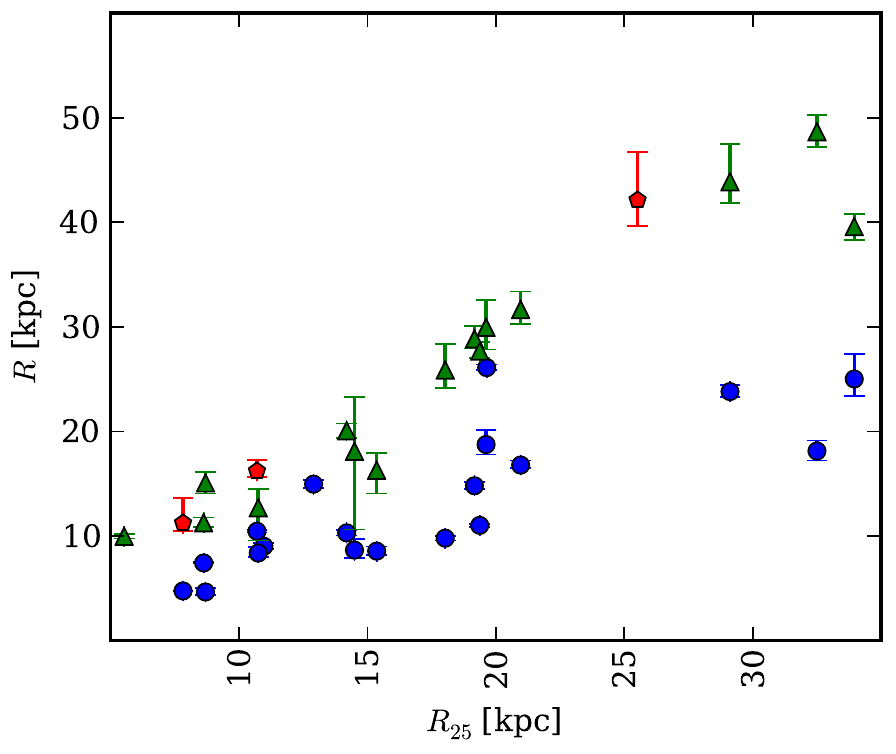}
%
% If no graphics program available, insert a blank space i.e. use
%\picplace{5cm}{2cm} % Give the correct figure height and width in cm

\caption{Feature size $R$ as a function of galaxy size, indicated by $R_{25}$, for the moderately inclined to
face-on galaxies in the sample of \citet{peters:2017}. Red pentagons represent truncations, green triangles the
onset of the stellar halo, and blue circles disk breaks which occur further inside the disk. Reproduced with
permission from \citet{peters:2017}}

\label{peters}       % Give a unique label
\end{figure}

Figure~\ref{peters} illustrates many of the results of \citet{peters:2017}. Firstly, disk breaks at relatively
high surface brightness were found in the radial profiles of most galaxies (blue dots in the figure), at levels of
typically $22-24\,r'$-mag\,arcsec$^{-2}$. Their radius scales with galaxy radius. Secondly, truncations were
indeed identified in three of these moderately inclined galaxies, at surface brightness levels of around
$28\,r'$-mag\,arcsec$^{-2}$. Thirdly,  in most galaxies a flattening of the radial profile is seen, interpreted as
the onset of the stellar\index{stellar haloes} halo\footnote{\citet{peters:2017} perform a careful PSF modelling
to ensure that the observed haloes are not artifacts caused by the PSF.}, and starting to dominate the profiles at
$28\pm1\,r'$-mag\,arcsec$^{-2}$. Fourthly, not only do the radii of the onsets of the truncation and the halo
components scale with galaxy size, for some reason they do that in the same manner, so that truncation and halo
points (red pentagons and green triangles, respectively, in Fig.~\ref{peters}) line up on the same relation. This
intringuing relation deserves further attention with other independent surveys, as these features are barely above
the surface brightness limit of the Stripe82.

A further very interesting observation from \citet{peters:2017} is that truncations are only observed in those
galaxies from which the halo component is either absent or fainter than usual, although no indications were found
from galaxy or feature parameters as to why the halo might be faint or absent. Of the seven galaxies with faint or
absent haloes, three were found to host truncations. This is not out of line with the statistical result of
\citet{kregel:2002} that at least 20 of their 34 edge-on galaxies were truncated. More detailed study of larger
samples of galaxies should shed further light on these important issues, and future developments are promising in
this respect (see Sect.~\ref{sec:5}).

\subsection{Tidal Streams}
\label{sec:4tidal}

Tidal streams\index{tidal streams} are the imprints of the ongoing merging activity\index{mergers of galaxies} of galaxies. They are present around
a significant fraction of galaxies, showing different shapes, brightnesses and extensions depending on their progenitor's mass
and orbit. The structural (and kinematical) properties of tidal streams can be used to infer 
information about the dark matter haloes\index{dark matter haloes} in which the host galaxies are embedded (\citealt{bullock:2005}). These
authors conclude that the vast majority of stream features have very low surface brightness
($\mu_V$$\gtrsim$28.5\,mag\,arcsec$^{-2}$). More precisely, cosmological simulations predict around one
detectable stream per galaxy if the observations reach a depth of $\sim$30\,mag\,arcsec$^{-2}$
(e.g., \citealt{bullock:2005,johnston:2008,cooper:2010}). This makes the use of very deep surveys mandatory if one wants to
have a complete census of past merging activity.

The detailed analysis of our Milky Way and the Andromeda galaxy M31 (using the star count technique) has revealed
a plethora of substructure in these objects. Due to the vicinity of these galaxies, the star count technique can
reveal features which have surface brightnesses equivalent to $\mu_V$$\sim$32\,mag\,arcsec$^{-2}$ (e.g.,
\citealt{ibata:2001,majewski:2003,belokurov:2006,bell:2008,McConnachie:2009}). These levels  are extremely
challenging to reach using integrated photometry (the only case published to date is that of the galaxy explored
in \citealt{trujillo:2016}). The need, however, for conducting an extensive survey exploring the variety of 
features around galaxies like the Milky Way is pressing. In fact, according to theory, a large galaxy-to-galaxy
variation is expected due to the intrinsic stochasticity of the merger phenomenon. For this reason, to have a
proper comparison with theory it is urgent to have a large survey exploring hundreds of galaxies beyond the Local
Group\index{Local Group}. This is only achievable using the integrated photometry technique. The star count
technique becomes unfeasible (with current telescopes) beyond $\sim$16\,Mpc (\citealt{Zackrisson:2012}).

The first attempt to systematize the search for streams in galaxies beyond the Local Group has been made by
\citet{martinez-delgado:2010}. These authors use modest ($\lesssim$1\,m) aperture telescopes using integration
times $\sim$10-20\,h, allowing them to reach $\mu_g$$\sim$28.5\,mag/arcsec$^2$. They have found a large variety
of tidal stream morphologies in eight nearby galaxies: large arc-like features, giant ``umbrellas", shells,
plumes, spiral-like patterns, etc. These morphologies are similar, at least qualitatively, to those expected in
cosmological simulations. However, the galaxies explored in \citet{martinez-delgado:2010}
are not represenative of the general galaxy population as they were preselected to have bright structure
surrounding them. In this sense, despite the effort of these authors, the comparison of theory with observations
remains currently in its infancy. This is because a large survey, probing hundreds of randomly selected galaxies, is still missing. 

\subsection{Stellar Haloes}
\label{sec:4halo}

Tidal streams and stellar haloes\index{stellar haloes} around galaxies are intimately connected. In fact, in observations with enough
resolution and signal-to-noise stellar haloes should appear as a combination of many different
merger events (and consequently, the sum of many tidal streams\index{tidal streams}). Stellar haloes also have a diffuse
component which corresponds to those accretion events that happened very early on in a galaxy's evolution. Whereas bright
tidal streams inform us about the ongoing or recent merging activity of a galaxy, stellar haloes give us
information about the past accretion story of galaxies. 

In contrast with numerical simulations, where the characterization of the stellar halo is relatively
simple, there is no agreement in the literature on how to quantify the properties of stellar haloes
using images of galaxies alone. Simulations (e.g., \citealt{cooper:2010,font:2011,tissera:2014,pillepich:2014}) show unequivocally that the
contribution of accreted stars rises towards the centre of galaxies. In fact, in numerical
simulations the accreted material follows surface mass distributions that can be modelled well with
S\'ersic (with S\'ersic indexes $n\lesssim3$) or power-law profiles (with logarithmic slope $\sim3.5$) for
the 3D density profiles. Accreted stars are the main contributors to the stellar mass surface
distribution of galaxies in their outer regions, however, towards the inner zones they contribute much less in comparison with the stars
that have formed {\it in situ} (a factor of $\sim$100 in projected density; \citealt{cooper:2013}).

\begin{figure}[htb]
%\sidecaption
% For example, with the graphicx style use
\includegraphics[width=\textwidth]{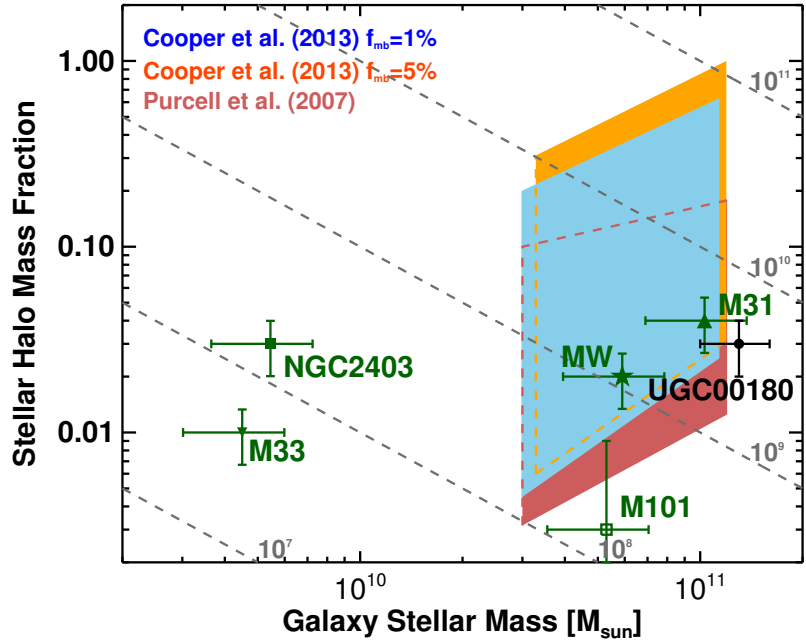}
%
% If no graphics program available, insert a blank space i.e. use
%\picplace{5cm}{2cm} % Give the correct figure height and width in cm

\caption{Stellar halo mass fraction versus total galaxy stellar mass, illustrating the
location of some nearby galaxies in this plane. In addition, the prediction from several numerical
models are overplotted (\citealt{purcell:2007,cooper:2013}). The dashed lines are the locations of stellar haloes with fixed
stellar mass (10$^7$, 10$^8$, 10$^9$, 10$^{10}$ and 10$^{11}\,M_\odot$). Figure from
\citet{trujillo:2016}, reproduced by permission of the AAS}

\label{fig:2}       % Give a unique label
\end{figure}

The observational characterization of stellar halo profiles has followed different approaches. Several
authors fit the outer (stellar halo-dominated) surface brightness region with some analytical profile,
for instance, exponential (e.g., \citealt{irwin:2005,ibata:2007,trujillo:2016}) or power-law (with
logarithmic slopes $\sim$2.5; e.g., \citealt{tanaka:2010,courteau:2011,gilbert:2012}). This approach has the
disadvantage of assuming a particular shape of the stellar
halo towards the inner region of galaxies, motivated by numerical simulations. This shape is derived by extrapolating the fit to the
outer region towards the inner area. There is no way of eliminating this assumption observationally with
integrated photometry alone, because in the central part of the galaxy the light is dominated by
the bulge and disk. Other authors avoid making any assumption about the inner shape of the stellar
halo profile and measure the properties of the stellar halo only in the outer regions where the halo is the
dominant contribution. For instance, \citet{buitrago:2016}, for elliptical galaxies, use the radial
region $10<R{\rm (kpc)}<50$ whereas \citet{merritt:2016} use $R{\rm (kpc)}>5\,R_{\rm h}$ (with $R_{\rm h}$ the half-mass
radius) for spiral galaxies. This approach has the disadvantage of neglecting the (dominant)
contribution of the stellar light of the halo in the innermost region of the galaxy. Any of the two
above approaches to measure the relevance of the stellar halo can be followed if a proper comparison
(i.e., following the same scheme) is conducted with the numerically simulated galaxies as well. In the
literature it is common to find a comparison of the amount of stellar mass in the stellar halo with the
total stellar mass of the galaxy. This relation is then compared with numerical predictions (see,
e.g., Fig. \ref{fig:2}). Observers directly measure neither the light contribution from the stellar haloes nor the
stellar mass. To transform from light to stellar mass it is necessary to estimate a stellar mass to light
ratio. This can be done using photometric information from several filters (when available).

The fraction of mass in the stellar haloes of disk (Milky Way-like) galaxies is around 1\%. However, there are notable
exceptions like M101, which does not show evidence for a stellar halo (\citealt{vandokkum:2014}) even when using deep
imaging. Observed stellar haloes are, in general, less massive than expected from theory. Nonetheless, the number
of observed galaxies is still too limited to consider this as a serious concern.

Measuring the amount of stellar mass in the stellar halo is a strong test to probe the predictions from
the $\rm \Lambda$-Cold Dark Matter ($\rm \Lambda$CDM)\index{$\rm \Lambda$CDM cosmology} model in the context of galaxy formation. Ultimately, the amount of stars in this component of
the galaxy informs us about the merging activity along the full history of the galaxy. Consequently, if
the observed stellar haloes are less massive than the theoretical expectations this might indicate that
the number of accretion events was less than presumed. If this were in fact the case, we could be
witnessing a problem for the $\rm \Lambda$CDM scheme as relevant as the missing satellite problem (see Sect.~\ref{sec:4sat}).

\subsubsection{PSF Effects on Imaging the Outermost Regions of Galaxies}
\label{sec:4halo_psf}

\begin{figure}[htb]
%\sidecaption
% For example, with the graphicx style use
\includegraphics[width=\textwidth]{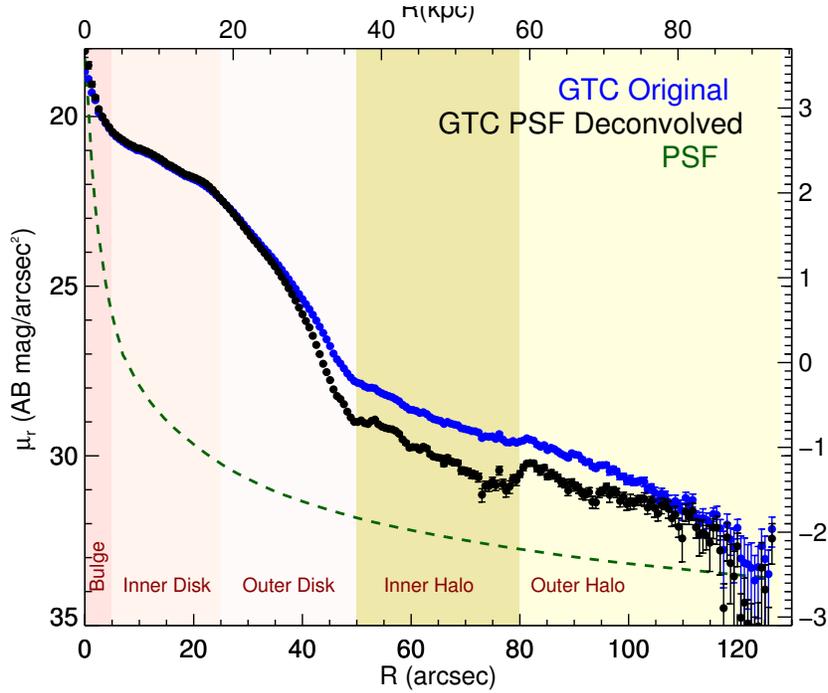}
%
% If no graphics program available, insert a blank space i.e. use
%\picplace{5cm}{2cm} % Give the correct figure height and width in cm

\caption{The effect of the PSF on the surface brightness profile of the spiral galaxy UGC~00180. The
observed profile (upper one) is shown using blue dots. The profile corrected for the effect of the PSF (lower profile) is illustrated
with black dots. The green dashed line is the surface brightness profile of the PSF of the image. The
different parts of the galaxy are indicated with labels. Figure from \citet{trujillo:2016},
reproduced by permission of the AAS}

\label{fig:3}       % Give a unique label
\end{figure}

As discussed in Sect.~\ref{sec:2scat}, one of the most important sources of light contamination
in deep images is the contribution of scattered light from nearby sources surrounding the
target galaxy. However, the galaxy itself is usually one of the most important sources of light contamination in
the vicinity of the object. This is easy to understand. The surface brightness distribution of the
galaxy is convolved with the PSF\index{point spread function} of the image. That means that a substantial amount of light coming from
the central parts of the galaxy is distributed into its outermost region.
\citet{sandin:2014,sandin:2015} has nicely illustrated this phenomenon, showing that
the scattered light from the object (if not accounted for) can be wrongly interpreted as a bright
stellar halo or a thick disk. This was also explored by \citet{dejong:2008}. Depending on the shape of the
surface brightness distribution of the galaxy, the effect of its own scattered light can be more or less
relevant in its outer region. Edge-on disks are normally most severely affected by this effect,
whereas face-on galaxies without breaks or truncations are barely affected. As a general rule, the
effect will be stronger in those cases where the light distribution is sharper. In Fig. \ref{fig:3}, we
show an example of the effect of the PSF on the light distribution of a spiral galaxy.

As Fig. \ref{fig:3} illustrates, the effect of the PSF on the outer regions of the galaxy is dramatic.
For this particular galaxy, the surface brightness profiles start to deviate at around 26\,mag\,arcsec$^{-2}$. At $\mu_r\sim$26\,mag\,arcsec$^{-2}$, the effect is so important that the surface brightness
profiles (affected by the PSF and corrected) are different by about 1\,mag. This means that in the outer
regions the scattered light can be as much as three times brighter than the intrinsic light of the stellar halo.
For this reason, it is absolutely necessary to model the effect of the PSF on the galaxy itself if one
wants to explore the faintest regions of the galaxies.

\begin{figure}[htb]
%\sidecaption
% For example, with the graphicx style use
\includegraphics[width=\textwidth]{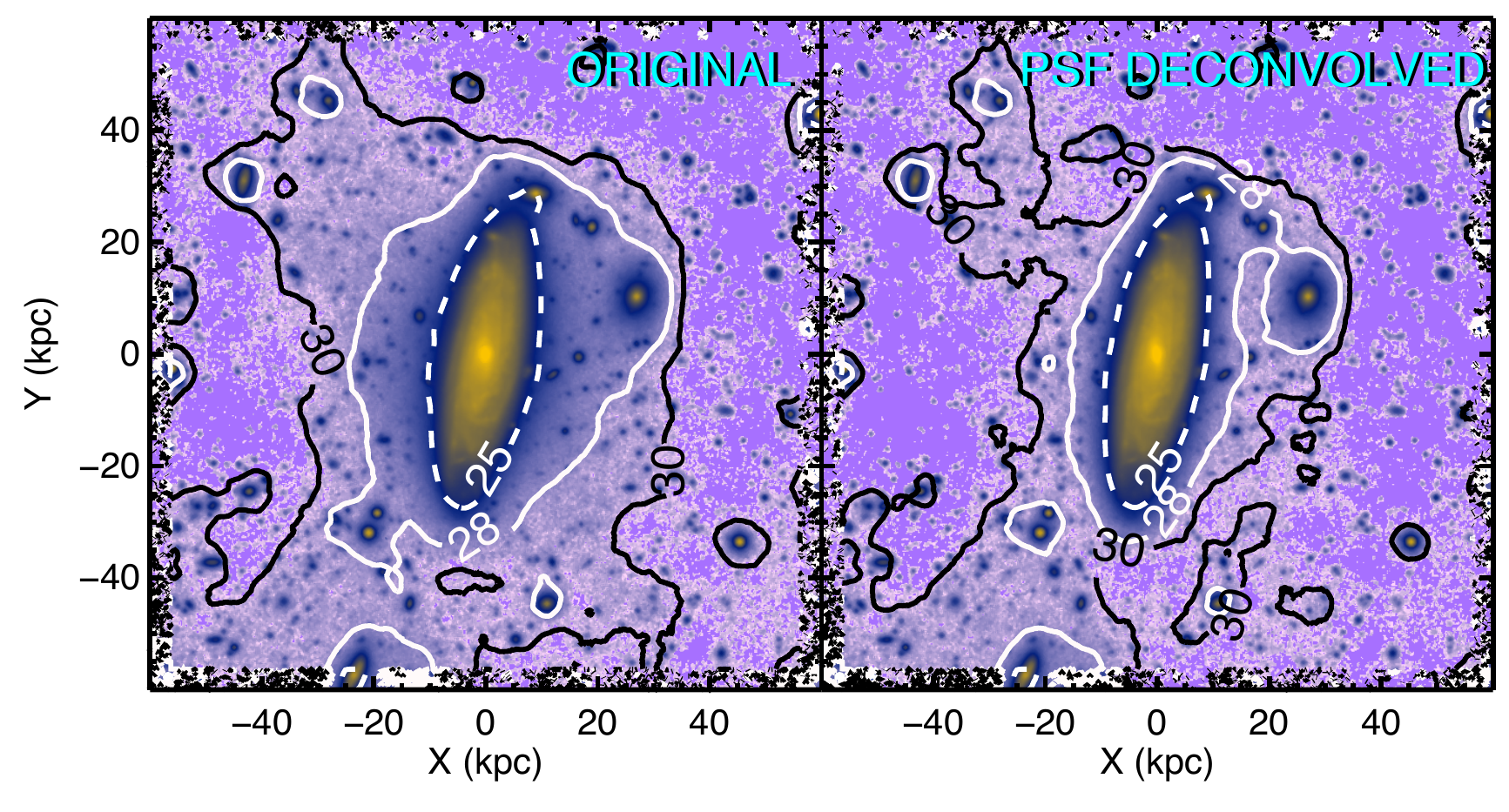}
%
% If no graphics program available, insert a blank space i.e. use
%\picplace{5cm}{2cm} % Give the correct figure height and width in cm

\caption{The dramatic effect of the PSF on the surface brightness distribution of the spiral galaxy
UGC~00180. Different surface brightness isophotes are indicated with solid and dashed lines. After
accounting for the effect of the PSF the amount of light in the outer regions decreases strongly. Figure
from \citet[][]{trujillo:2016}, reproduced by permission of the AAS}

\label{fig:4}       % Give a unique label
\end{figure}

There are two different approaches to handle the effect of the PSF on the surface brightness
distribution of the galaxies. The first one is to apply a deconvolution method directly on the images.
This has the advantage that it only requires an exquisitely characterized PSF of the image. However, its
main disadvantage is that in the regions where the surface brightness is closest to the noise the result is quite uncertain. The second
approach is to model the intrinsic light distribution of the galaxy and convolve this with the PSF.
The convolved model of the galaxy is then fitted to the observed light distribution until a good fit to
the data is reached. Once this is achieved, a new image of the object (using the deconvolved model
of the galaxy) is created, adding back the residuals of the fit. This is illustrated in Fig.~\ref{fig:4}.

Figure~\ref{fig:4} shows that studies of both thick disks and outer stellar haloes are strongly
affected by the PSF. Any deep imaging study of these components needs to account for this effect,
otherwise both the stellar haloes and the thick disks that will be inferred will result significantly brighter
than what they really are.

\subsection{Satellites}
\label{sec:4sat}

Satellites\index{satellite galaxies} are probably one of the most important pieces to connect the realm of obseved galaxies with 
cosmological modelling. In fact, counting the number of satellites and exploring their properties has
become one of the most important tests to explore the predictions of the $\rm \Lambda$CDM\index{$\rm \Lambda$CDM cosmology} model on small
(i.e., galactic) scales. The analysis of Local Group\index{Local Group} galaxies has revealed a number of problems
related with satellite galaxies. These can be summarized as follows:

\begin{itemize}

\item The ``missing satellites" problem. Cosmological simulations, in the framework of the
$\rm \Lambda$CDM model and based only on dark matter particles (i.e., without including the effect of 
baryons), predict thousands of small dark matter haloes orbiting galaxies like the Milky Way or Andromeda
(e.g., \citealt{klypin:1999,moore:1999}). However, only a few tens of visible satellites have been
observed around these objects (\citealt{mcconnachie:2012}).

\item The ``too big to fail" problem. In the $\rm \Lambda$CDM model, the largest satellite galaxies are
expected to have a velocity dispersion substantially larger than what is actually measured in any of the
dwarf galaxies in the Local Group (\citealt{read:2006,Boylan-Kolchin:2011}). These big haloes among the satellite population should
not have failed to produce dwarf galaxies following the prescription of baryonic physics, so if they exist, dwarf galaxies
with such large central velocity dispersion should have been found observationally.  

\end{itemize}

There are a number of solutions to the above cosmological problems at galactic scales. One
possibility is that the main constituent of dark matter would not be a heavy (i.e., cold) particle
but instead a lighter  (i.e., warm) candidate (with a mass of around a keV). This hypothesis has been
explored in many papers as it is able to suppress many of the dark matter haloes\index{dark matter haloes} which should form dwarf
galaxies\index{dwarf galaxies} (e.g., \citealt{moore:1999,bode:2001,avila-reese:2001}). This alleviates the problem of the
missing satellites. This solution is not free of other problems, however, mainly the current constraint
on the mass of the warm dark matter candidate imposed by the study of the Lyman $\alpha$ forest
(i.e., $\gtrsim 4$\,keV; \citealt{baur:2016}). This mass is large enough to mimic most of the properties of
the  $\rm \Lambda$CDM scenario (including the production of a large number of dark matter subhaloes).
Consequently, warm dark matter particle candidates with such minimum mass would not be of great help.
Another alternative is to consider the baryon physics in detail when creating realistic cosmological
simulations. It is worth stressing that the above problems arise when comparing simulations of dark
matter alone (i.e., devoid of gas and stars) with the frequency and properties of satellites in the Local
Group. However, some authors, like \citet{navarro:1996} or \citet{read:2005}  have pointed out the interplay
between baryons and dark matter. In fact, they have suggested that the dark matter could be
heated by impulsive gas mass loss driven by supernova explosions. If so, this could help to reconcile the
cosmological simulations with the observations.

The above discrepancies between theory and observations are based on the comparison of the
properties of the satellites galaxies of two massive disks, i.e., those of our own Galaxy and Andromeda.
Consequently, a natural question arises: is our Local Group anomalous in producing a low number of
satellites? To address this question in detail, we need to characterize the satellite population of many
galaxies like ours beyond the Local Group. This idea has been pursued by several authors (see,
e.g., \citealt{liu:2011,ruiz:2015}). Unfortunately, their analyses have only explored the population of
the most massive satellites (i.e., down to masses $\sim10^8-10^9\,M_\odot$). To test the
discrepancies between the cosmological models and the observations it is necessary to go at least two orders
of magnitudes fainter (i.e., down to $10^6\,M_\odot$). This is where new ultra-deep imaging surveys
can play a definitive role (see, e.g., \citealt{javanmardi:2016}). To explore satellites with masses around $10^6\,M_\odot$ within a sphere of
radius $\sim100$\,Mpc, we need to be able to reach a depth like the one achieved by Stripe~82. If we want to
go further and explore a population of satellites with masses of a few times $10^4\,M_\odot$ (which is
currently the mass limit of detected satellites in Andromeda) we would need a survey with the depth
reached by the current deepest dataset (i.e., the one of \citealt{trujillo:2016}). A survey with such characteristics
could be obtained at the end of the LSST programme (\citealt{ivezic:2008}; see Sect.~\ref{sec:5}). 

\section{Conclusions and Future Developments}
\label{sec:5}

In this Chapter, we have reviewed how deep imaging is a fundamental tool in the study of the outermost structure
of galaxies. Three main sources of imaging are currently used to detect and characterize the outskirts of
galaxies: (1) surveys such as the Sloan Digital Sky Survey's Stripe~82 project, (2) very long exposures on small
telescopes, including by amateurs, and (3) long exposures on the largest professional telescopes. The technical
challenges in overcoming systematic effects are significant, and range from the treatment of light scattered by
the atmosphere and the telescope and instrument, via flat fielding, to the accurate subtraction of non-galaxy
light in the images. We have reviewed recent results on galaxy disks and haloes obtained with deep imaging,
including the detection and characterization of thick disks, truncations of stellar disks, tidal streams, stellar
haloes, and satellites. We have shown how each of these interrelated aspects of the faintest detectable structure
in galaxies can shed light on the formation and subsequent evolution of the galaxies and our Universe. 

The future is promising in terms of discovering the ``low surface brightness Universe" through deep imaging.
Current techniques using small and large ground-based telescopes, as reviewed in Sect.~\ref{sec:3}, will continue
to be exploited while new facilities will become available. The most promising of these new facilities are poised to be the
ground-based LSST, and the new space telescopes {\it JWST} and {\it Euclid}. LSST\index{LSST}
(\citealt{ivezic:2008}) will repeatedly image the whole of the Southern sky in 6 passbands ($ugrizy$), with a
planned start date for surveys of around 2021. While each 30\,s exposure with the 8.2\,m telescope and optimized
camera will yield a depth of 24.7\,$r$-mag (5-sigma point source depth) over the 9.6\,square degree field of view,
combining all imaging obtained over the approximately 10\,year survey duration could yield a depth of
27.5\,$r$-mag (5-sigma point source depth). This roughly means as deep as the current Stripe~82 in one exposure,
and close to three magnitudes deeper over the whole survey duration. The scientific possibilities offered by this
depth of imaging, over an area of sky of over 20000\,deg$^2$, are tremendous. All the science discussed in this
paper could essentially be done with one or a few exposures, or easily with the data of say one hour of
observation. This obviously depends critically on whether systematic effects, including but not limited to those
discussed in Sect.~\ref{sec:2}, can be properly controlled, modelled, and corrected for. This is not trivial, and
may hinder the full exploitation of the data to their theoretical limits.

The {\it JWST}\index{JWST}, to be launched in 2018, will allow deep imaging but, when compared to, e.g., the {\it HST}, won't be quite as revolutionary as LSST. The field of view will be limited to just over $2\times4$\,arcmin$^2$ in the
near-IR which will all but exclude deep imaging of the nearest galaxies. Where significant progress can be
expected is in the deep imaging of galaxies at redshifts of, say, 0.2 and higher. In particular, near-IR imaging
will allow the observation of galaxies at redshifts beyond 1 in rest-frame red passbands, necessary to reduce the
effects of both young stellar populations and dust extinction. 

{\it Euclid}\index{Euclid} will provide, among other data products, imaging in a very wide optical band ($R+I+Z$)
over an area of 15000\,deg$^2$ to a depth of 24.5\,mag (10$\sigma$ for extended source). Compared to LSST,
advantages of {\it Euclid} imaging will be its higher spatial resolution (of $\sim$0.2 arcsec, due to the
relatively small telescope aperture of 1.2\,m) and better-behaved PSF, thanks to the absence of the Earth
atmosphere. A disadvantage is that the visual imaging is done through a very wide filter which excludes the use of
colour information. The depth of imaging will be comparable with LSST, though, for {\it Euclid}'s Deep Survey (over an area of around 40
square degree), which will allow the comparison of high-resolution {\it Euclid} imaging with the colour
information obtained from the LSST images. 

The area of ultra-deep imaging is still very much unexplored territory, and future work in this area will be
stimulated by the availability of revolutionary new data sets and the continued understanding of the systematics
affecting them. It will be a huge technical challenge to properly treat and analyze the upcoming deep imaging, in
particular those from LSST and {\it Euclid}. As we enter previously unexplored territory with ultra-deep imaging,
the systematic effects we know about will be challenging to characterize and correct for, and additional
difficulties will almost certainly present themselves. But overcoming these issues will definitely pay off in
terms of increased understanding of the formation and evolution processes which have led to the Universe and the
galaxies as we observe them now. The future of imaging ultra-faint structures is very bright.

\begin{acknowledgement}

JHK thanks S\'ebastien Comer\'on and Carme Gallart for comments on sections of the manuscript. IT has benefitted
from multiple conversations with and the hard work of many members of his team. In particular, he thanks J\"urgen
Fliri, Mar\'ia Cebri\'an and Javier Rom\'an. JHK and IT acknowledge financial support from the Spanish Ministry of
Economy and Competitiveness (MINECO) under grants number AYA2013-41243-P and AYA2013-48226-C3-1-P, respectively. JHK acknowledges financial support to the DAGAL network from the People Programme (Marie Curie Actions) of the European UnionÕs Seventh Framework Programme FP7/2007-2013/ under REA grant agreement number PITN-GA-2011-289313.

\end{acknowledgement}

\bibliographystyle{spbasic}

\bibliography{KnapenTrujillo_references}

%\printindex

%\input{KnapenTrujillo_referenc}
\end{document}